\documentstyle[12pt,amssymb,amscd]{article}

\def\AFOUR{%
\setlength{\textheight}{9.0in}%
\setlength{\textwidth}{5.75in}%
\setlength{\topmargin}{-0.375in}%
\hoffset=-.5in%
\renewcommand{\baselinestretch}{1.17}%
\setlength{\parskip}{6pt plus 2pt}%
}
\AFOUR
\def\car{\mathop{\square}}
\def\carre#1#2{\raise 2pt\hbox{$\scriptstyle #1$}\car_{#2}}

\makeatletter
\def\section{\@startsection {section}{1}{\z@}{-3.5ex plus -1ex minus
 -.2ex}{2.3ex plus .2ex}{\large\bf}}
\def\subsection{\@startsection{subsection}{2}{\z@}{-3.25ex plus -1ex minus
 -.2ex}{1.5ex plus .2ex}{\normalsize\bf}}
\makeatother
\makeatletter
\@addtoreset{equation}{section}

\makeatother
\newcommand{\nc}{\newcommand}
\newcommand{\rnc}{\renewcommand}
\nc{\be}{\begin{equation}}
\nc{\ee}{\end{equation}}
\nc{\ba}{\begin{eqnarray}}
\nc{\ea}{\end{eqnarray}}

\def\slash#1{\setbox0=\hbox{$#1$}#1\hskip-\wd0\hbox to\wd0{\hss\sl/\/\hss}}

\def\href#1#2{{#2}}

\rnc{\a}{\alpha}
\nc{\ab}{\bar{\a}}
\nc{\ap}{\a^{+}}
\nc{\abm}{\ab^{-}}
\rnc{\b}{\beta}
\nc{\bb}{\bar{\b}}
\nc{\bbp}{\bb_{\zb}^{+}}
\nc{\bm}{\b_{z}^{-}}
\nc{\oa}{\overline{\a}}
\nc{\ob}{\overline{\b}}
\rnc{\gg}{\gamma}
\rnc{\d}{\delta}
\nc{\f}{\phi}
\nc{\fb}{\bar{\phi}}
\nc{\vf}{\varphi}
\nc{\p}{\psi}

\rnc{\c}{\chi}
\nc{\la}{\lambda}
\nc{\m}{{\mathrm m}}
\nc{\n}{\nu}
\rnc{\o}{\omega}
\nc{\Om}{\Omega}
\rnc{\t}{\theta}
\nc{\eps}{\epsilon}
\rnc{\S}{\Sigma}
\nc{\F}{\Phi}
\nc{\trac}[2]{{\textstyle\frac{#1}{#2}}}
\nc{\ex}[1]{\mbox{e}^{\,\textstyle#1}}
\nc{\mat}[4]{\left(\begin{array}{cc}#1&#2\\#3&#4\end{array}\right)}
\nc{\som}[9]{\left(\begin{array}{ccc}#1&#2&#3\\#4&#5&#6\\#7&#8&#9%
\end{array}\right)}
\nc{\tr}{\mathop{\mbox{tr}}\nolimits}
\nc{\ad}{\mathop{\mbox{ad}}\nolimits}
\nc{\Tr}{\mathop{\mbox{Tr}}\nolimits}
\nc{\Det}{\mathop{\mbox{Det}}\nolimits}
\nc{\rk}{\mathop{\mbox{rk}}\nolimits}
\nc{\ra}{\rightarrow}
\nc{\Ra}{\Rightarrow}
\nc{\LRa}{\Leftrightarrow}
\nc{\ot}{\otimes}
\rnc{\ss}{\subset}
\nc{\nul}{\noindent\underline}
\nc{\non}{\nonumber\\}
\nc{\subs}[1]{{\vspace*{0.5cm}}%
{\noindent\underline{#1}}{\addcontentsline{toc}{subsection}{#1}}%
{\vspace*{0.3cm}}}
\nc{\zb}{\bar{z}}
\rnc{\lg}{\frak{g}}
\nc{\lt}{\frak{t}}
\nc{\lk}{\frak{k}}
\nc{\lh}{\frak{h}}
\nc{\pik}{\Pi_{\lk}}
\nc{\pip}{\Pi_{+}}
\nc{\pim}{\Pi_{-}}
\nc{\pih}{\Pi_{\lh}}
\nc{\jz}{J_{z}}
\nc{\jzh}{\jz^{\lh}}
\nc{\jzp}{\jz^{+}}
\nc{\jzm}{\jz^{-}}
\nc{\del}{\partial}
\nc{\dz}{\del_{z}}
\nc{\dzb}{\del_{\bar{z}}}
\nc{\az}{A_{z}}
\nc{\azb}{A_{\bar{z}}}
\nc{\g}{g^{-1}}
\nc{\dw}{\Delta_{W}}
\nc{\Ad}{{\mbox{Ad}}}
\nc{\ks}{Ka\-za\-ma-\-Su\-zu\-ki}
\nc{\KS}{\ks}
\nc{\ksm}{\ks\ model}
\rnc{\AA}{{\Bbb A}}
\nc{\BB}{{\Bbb B}}
\nc{\CC}{{\Bbb C}}
\nc{\PP}{{\Bbb P}}
\nc{\cpm}{\CC\PP(m)}
\nc{\cpn}{\CC\PP(n)}
\nc{\cp}[1]{\CC\PP(#1)}
\nc{\gmn}{G(m,m+n)}
\nc{\gmnk}{\gmn_{k}}
\nc{\cO}{{\cal O}}
\nc{\bcO}{\bar{\cO}}
\nc{\bO}{\bar{O}}
\nc{\oQ}{\overline{Q}}
\nc{\ie}{{\it i.e.~}}
\nc{\eg}{{\it e.g.~}}
\begin{document}
\makeatother\begin{titlepage}
\begin{flushright}
{}
\end{flushright}
\vspace*{0.1in}
\begin{center}
{\Large\bf Anomalies, RG-flows and \\ Open/Closed String Duality }\\
\vskip .3in
\makeatletter
\centerline{ Massimo Bianchi \footnote{M.Bianchi@damtp.cam.ac.uk} and
Jose F. Morales
\footnote{morales@phys.uu.nl } }
\bigskip 
{\it D.A.M.T.P. University of Cambridge, UK \\
Dipartimento di Fisica and sez. I.N.F.N., \\
Universit{\`a} di Roma ``Tor Vergata'', Italia$^1$}

\smallskip
{\it Spinoza Institute, Utrecht, The Netherlands $^2$}


\end{center}

\vskip .10in

\begin{abstract}
{We discuss the interplay between IR and UV divergences in
vacuum configurations with open and unoriented strings. We establish a
general one-to-one correspondence between anomalies
and R-R tadpoles associated to sectors with non-trivial Witten index.
The result does not require any supersymmetry to be preserved by the
configuration. Under very mild conditions of supersymmetry,
a similar correspondence is found between NS-NS tadpoles
and RG-flows in gauge theories on D-branes and O-planes.
We briefly comment on the AdS/CFT counterpart of the results.}
\end{abstract}

\vfill
\noindent
{\it MG9 Proceedings: submitted to World Scientific on January 12,2001.}

\noindent
{\it Keywords: Anomalies, Tadpoles, RG-flows, Open/Closed string duality }

\end{titlepage}

\section{Introduction}
The interplay between gauge theories and (super)gravity
is one of the central issue in the string theory.
In the past few years, some remarkable forms of duality
between the two have been conjectured
by BFSS \cite{bfss} and Maldacena
\cite{maldacena}.
In our view, the duality between open and closed
string channels offers the cleanest way to understanding the
relation between gauge theories governing
the low energy dynamics of configurations of
D-branes and O-planes and certain (super)gravity theories living in the
bulk of spacetime. A generic string amplitude around such
backgrounds can be either thought in terms of closed
string exchange between boundary and crosscap states or in terms
of open string loops.
In the limits where the worldsheet degenerates into long closed-string
tubes or into thin open-string ribbons, only massless states
contribute
in either description
and we can effectively interpret the results
in terms of gauge theory or of supergravity.
Of course, the two answers will in general disagree since they correspond
to rather different truncations of the complete result.
Under special conditions, however, they can be shown to coincide.
In \cite{mss} these ideas were exploited in the context of
Matrix theory. The leading order spin interactions
between slowly moving parallel D-branes were extracted from
fullfledged one-loop (in the open string channel) or tree level (in the
closed string channel) amplitudes. The same result was shown to
admit equivalent super Yang-Mills and
supergravity descriptions.
Aim of this talk is to review the correspondence between
one-loop IR divergences, such as those giving rise to anomalies
and RG-flows in gauge theories on D-branes and O-planes, and UV divergences
associated to tadpoles in their dual supergravity descriptions.
A complete exposition and references can be found in \cite{mm1,mm2}.
We follow the notations and techniques developed in \cite{torvergata}.
We will denote by ${\cal K}$, ${\cal A}$ and ${\cal M}$, respectively,
the  contributions of Klein bottle, Annulus and Moebius strip to the
relevant string amplitudes. By $\tilde{\cal K}$,
$\tilde{\cal A}$ and $\tilde{\cal M}$ we denote similar expressions
rewritten in the closed string channel.
The two sets are mapped into one another by
(model dependent)
modular transformations that exchange
$\sigma$ and $\tau$ directions on the worldsheet.
At first sight one may thus in general doubt the existence of any connection
between the truncations to their massless contributions of
either description.
A carefull look into the structure of certain string amplitudes reveals
that this is not always the case. In some cases, much as in the Matrix
model computations, the zero-mode contribution happens to be exact and
both descriptions turn out to be accurate.

\section{Anomalies and R-R tadpoles}
Let us consider a generic vacuum configuration containing
open and unoriented strings. It can be specified in terms of its
one-loop partition function\footnote{We will always omit
the modular invariant torus contribution of
the parent closed-string theory. Being anomaly free it is
irrelevant for our present purposes.} ($\int\frac{dt}{t}$)
\ba
{\cal K}&=&{1\over 2\,t^{D\over 2}} K^{i}{\cal X}_{i}(2it)\nonumber\\
{\cal A} &=& {1\over 2\, t^{D\over 2}} A^{i}_{ab} n^{a} n^{b}
{\cal X}_{i} \left( {it\over 2}\right) \nonumber\\
{\cal M}&=&{1\over 2\,t^{D\over 2}}  M^{i}_{a} n^{a}  \widehat{\cal X}_{i}
\left( {it\over 2} + {1\over 2}\right)
\label{direct}
\ea
with ${\cal X}_{i}$ a basis of charaters in the SCFT,
$a,b$ running over the number of boundaries,
\ie independent Chan-Paton charges, with integer multiplicities $n^a$.
The integers $K^{i}, A^{i}_{ab}$, $M^{i}_{a}$
count the number of times the sector $i$ flows in the
Klein bottle, Annulus and M\"obius-strip loop, respectively
\footnote{Here and in the following, we will not
distinguish between ``complex'' (unitary) and ``real''
(orthogonal or symplectic) Chan-Paton multiplicities $n^a$.
The sum over $a$ will include
two contributions for the former and only one for the latter, thus
producing not only the correct dimensions of the representations but
also the correct orientation of the boundary.}.
In terms of closed string variables one can alternatively
write ($\int dq$)
\ba
\widetilde{\cal K} &=& {2^{D\over 2}\over 2}
\sum_{i} (\Gamma^{i})^{2} {\cal X}_{i}(q)
\nonumber\\
\widetilde{\cal A} &=& {2^{-{D\over 2}}\over 2}
\sum_{i,a} (B^{i}_{a}n^{a})^{2}{\cal X} _{i}(q) \nonumber\\
\widetilde{\cal M} &=& {2\over 2}
\sum_{i,a} ( \Gamma^{i} B^{i}_{a}n^{a}) \widehat{\cal X}_{i}(-q)
\label{transverse}
\ea
The relative powers of 2 result from the different rescalings of the modular
parameters ($\tau_{{\cal K}}= 2it, \tau_{{\cal A}} = it/2,
\tau_{{\cal M}} = it/2 + 1/2$)
that naturally enter the definition of the amplitudes in
the direct channel. These rescalings are necessary in order for the
amplitudes in the transverse channel to be expressed in terms of the common
length of the tube $\ell = - {1\over 2\pi} \log {q}$.
The coefficients $\Gamma^{i}$ and $B^{i}_{a}$ should then be
interpreted as the reflection coefficient (one-point function)
on a crosscap and on a boundary of type $a$, respectively.
They are related to the integer coefficients in the direct channel
by suitable modular transformations:
\ba
K^{i} &=& \sum_{j} S^{i}{}_{j}\Gamma^{j}\Gamma^{j} \nonumber\\
A^{i}_{ab} &=& \sum_{j} S^{i}{}_{j}B^{j}_{a}B^{j}_{b} \nonumber\\
M^{i}_{a} &=& \sum_{j} P^{i}{}_{j}\Gamma^{j} B^{j}_{a}
\label{worldsheet}
\ea
with $S$ the modular matrix in the character basis
and $P\equiv T^{1/2} S T^2 S T^{1/2}$.

In $D$ (even) non-compact dimensions,
anomalies are associated to
one-loop string amplitudes in the odd-spin structure
involving a total number of $D/2 + 1$ graviton and/or gauge field
insertions (see \cite{ikk,ss,anomss,mm1} for details).
One of the vertices has to be taken with longitudinal polarization.
The results can be packaged into an anomaly generating polynomial,
whose three contributions read \cite{mm1}
\ba
{\cal K}_{odd} &=&
      -{1 \over 2}\sum_{i} {\cal I}_{i} K^{i}\,I_{A}(R)
\label{Kanomaly}\nonumber\\
{\cal A}_{odd} &=&
 {1 \over 4}\sum_{i,a,b}{\cal I}_{i} A^{i}_{ab} ch_{n^{a}}(F) ch_{n^{b}}(F)
 I_{{1/2}}(R)
            \nonumber\\
{\cal M}_{odd} &=&
    {1 \over 4}\sum_{i,a}{\cal I}_{i} M^{i}_{a} ch_{n^{a}}(2F)
          I_{{1/2}}(R)
            \label{Manomaly}
            \ea
Sum over repeated indices is always understood, unless differently
stated. $ch(F)$ is
the Chern character of the gauge fields. $I_{A}(R)$ and $I_{{1/2}}(R) $
represent the contributions to the gravitational anomaly
of a self-dual antisymmetric tensor
and a complex spin 1/2 L-fermion, respectively.
The important point is that since vertex insertions are aligned along the
spacetime directions the internal theory enters only through its
partition function in the odd spin structure denoted by
${\cal I}_{i}$ and known as the Witten index. Being topological
${\cal I}_{i}$ is an integer
independent of the parameters of the theory and in particular
of the worldsheet $t$-modulus.
The additional factor of one-half in the Annulus and M\"obius-strip
amplitudes reflects the fact that they are counting real fermions. The
relation between spin and statistics is responsible for the extra minus
sign of the Klein-bottle contribution with respect to the Annulus and
M\"obius strip. The former can only contribute loops of
(anti)self-dual antisymmetric tensors while the latter can only contribute
fermionic loops.

The consistency of the string background requires that irreducible
terms  ${\rm tr} R^{{D\over 2}+1}$, $tr_{n^a}\,F^{{D\over 2}+1}$
in the expansion of (\ref{Manomaly}) cancel.
Cancellations of gauge anomalies imposes a set of conditions
on the coefficients $A^{i}_{ab}$, $M^{i}_{a}$,
while the absence of gravitational anomalies leads to an
additional constraint on $K^{i}, A^{i}_{ab}$, $M^{i}_{a}$.
Saturating (\ref{worldsheet}) with ${\cal I}_i$ and using
modular invariance of the Witten index,
${\cal I}_{i} S^{i}{}_{j}= {\cal I}_{j}$ and
${\cal I}_{i} P^{i}{}_{j} ={\cal I}_{j}$, one can translate the conditions
arising from (\ref{Manomaly}) in terms of closed string variables.
After some manipulations, one can see that cancellation of gauge
anomalies requires
\be
{\cal I}_{i}\left(2^{D/2}\Gamma^{i}+B^{i}_{a} n^{a}\right)=0
\label{quasitadpole} \quad
\ee
where the index ``i'' is not summed over.
This precisely reproduces all tadpole cancellation conditions
for massless R-R closed string states belonging to
sectors with non-vanishing Witten index
${\cal I}_{i}\neq 0$. It is amusing to observe that
once this condition is satisfied  irreducible gravitational
anomalies are automatically cancelled, i.e.
in a consistent open string descendant
the absence of irreducible gauge anomalies
always implies the absence of irreducible gravitational
anomalies.
We conclude that: {\it There is a
one-to-one correspondence between the conditions for
cancellation of gauge anomalies and RR-tadpoles
associated to sectors of the internal SCFT
with non-vanishing Witten index.}
Tadpoles associated to
sectors in the SCFT with vanishing Witten index can be
related to ``higher dimensional anomalous amplitudes'' or
to gauge anomalies living on D-brane probes of the vaccum
geometry \cite{uranga}.
The early results \cite{cp,msa} and the systematic studies
\cite{bi,ibanez} represent important intermediate steps
to the very general final result.

\section{NS-NS tadpoles and RG-flows}
The correspondence between anomalies and RR tadpoles apply to any
consistent string vacuum configuration containing
open and unoriented strings, independently of the presence
of any unbroken spacetime supersymmetry.
In supersymmetric cases, we can go a step further and consider CP-even
couplings that still admits equivalent gauge/supergravity
descriptions.
Intuitively this should be clear. The sum over
even spin structures translates via the ``abstruse'' $\vartheta$-identity
into an odd spin structure contribution
to which our previous arguments apply.
Let us consider for example
${\cal N}=2$ supersymmetric gauge theories
in $D=4$ dimensions living on N D3-branes in the
presence of a background of D7-branes and O7-planes.
The running of the four dimensional
gauge coupling constant is logarithmic. The
coefficient can be extracted from a two-point
string amplitude involving two open strings ending
on the D3-branes \cite{bf,abd}.
The result can be written as \cite{mm2}
\be
\beta=
{1\over 4}(A^{i}_{ab} n^{b} +2 M^{i}_a)h_i=
{1\over 2}{\rm l}_2({\cal R}^{a}_i)h_i
\label{betad}
\ee
with $Tr_{{\cal R}_i} T^a T^b={\rm l}_2({\cal R}^{a}_i)\delta^{ab}$
and $h_i=-{11\over 3} n^i_V+{1\over 6}n^i_O+{4\over 3}n^i_F$
the standard field theory contribution to the $\beta$ function
of $n^i_O$ scalars, $n^i_V$ vectors and $n^i_F$ Dirac
fermions. As usual ``$i$'' labels the (massless)
characters of the spectrum.
In the closed string channel the logarithmic running is
associated to massless exchange in the two overall transverse
directions and the coefficient can be matched with (\ref{betad}).
Indeed noticing that $h^i$ coincides in ${\cal N}=2$ with ${\cal I}_i$
one can use (\ref{worldsheet}) and
${\cal I}_{i} S^{i}{}_{j}= {\cal I}_{j}$,
${\cal I}_{i} P^{i}{}_{j} ={\cal I}_{j}$ to translate
(\ref{betad}) into the closed string result
\be
\beta=2
(2^{-3}B^{i}_{b} n^{b} +{1\over 4}\Gamma^{i})B^{i}_a
h_i=2\hat{T}^{i} B^{i}_a h_i
\label{match}
\ee
with $\hat{T}^{i}=(2^{-3}B^{i}_{b} n^{b} +
{1\over 4}\Gamma^{i})$
the tadpole associated to the character ``$i$'' and localized
at one of the $4$ O7-planes. It is worth stressing that although
${\cal I}_i=h^i$ the two  types of divergences have quite different
origins. A careful look at the
linear combination of tadpoles appearing in (\ref{match}) reveals
that RG-flows is induced by NS-NS tadpoles, and more precisely
by the ten-dimensional dilaton tadpole, as expected.

Altohugh the above correspondence between RG-flow and NS-NS tadpoles
strongly relies on the presence of ${\cal N}=2$
supersymmetry, some extension is still possible.
One can consider D3-branes probing some
``F-theory'' backgrounds that admit a description in terms
of O7-planes and  D7-branes at angles. Since an
open string with one end on the D3-branes can at most have the other end
on one set of D7-branes, states contributing to the $\beta$-functions
organize themselves into supermultiplets of an ${\cal N}=2$
supersymmetry that is generically broken by the remaining sets of branes.
At least to leading order in 1/N, the correspondence is not expected
to be spoiled by higher loop corrections. The potentially
dangerous contributions come from worldsheets where
all except one of the boundaries lie on the D3-brane.
Again an underlying ${\cal N}=2$ supersymmetry prevents their onset.
In the following table we list several examples of ${\cal N}=1,2$
brane configurations where equivalent gauge/supergravity
descriptions of the $\beta$ function coefficients were discussed
in \cite{mm2}
\\
\begin{tabular}{lll}
Background  & Gauge Group & $\beta_N$ \\
D7& $U(N)\times U(n_A)\times U(n_B)$ & $n_A+n_B$\\
D7-O7 & $Sp(N)\times SO(n_A)$ & ${1\over 2}(n_A-8)$\\
$C^4/Z_2$ & $U(N)\times U(n_A)$ & ${1\over 2}(n_A+\bar{n}_A-8) $\\
$C^6/Z_2\times Z_2$ & $Sp(N)\times Sp(n_A)\times Sp(n_B)\times
Sp(n_C)$ &${1\over 2}(n_A+n_B+n_C-12)$
\end{tabular}
\\
From the table one can easily identify the values of $n$'s for which the
background is conformal, \ie $\beta_N=0$.

Finally we woulk like to briefly put the
above results in the perspective of the AdS/CFT correspondence.
The presence of a dilaton tadpole can be seen as a perturbation
of the supergravity field equations from the AdS configuration (conformal
point). Fortunately the perturbed equation
for the dilaton can be explicitly solved \cite{mm2}
in the large N limit and one is left with a logarithmically
running (in the radial coordinate).
The coefficient is precisely given by the
dilaton tadpole in agreement with our previous analysis.
Similar results for D3-branes probing a type 0 background
were originally found in \cite{aa}.

\section*{Acknowledgements}
We would like to thank the participants to the Meeting
for fruitful and stimulating discussions. This work is partly
supported by the PPARC grant, by the EEC contract HPRN-CT-2000-00122,
and by the INTAS project 991590.

\end{document}